\documentstyle[emulateapj]{article}

\newcommand{\gtilde}
 {~ \raisebox{-1ex}{$\stackrel{\textstyle >}{\sim}$} ~}
\newcommand{\ltilde}
 {~ \raisebox{-1ex}{$\stackrel{\textstyle <}{\sim}$} ~}

\begin{document}

\begin{flushright}
NAOJ-Th-Ap/00-24
\vspace*{-0.3cm}
\end{flushright}

\date{\today}
\submitted{Accepted by the Astrophysical Journal 
(Received 2000 May 9; Accepted August 3)}

\title{Forming Clusters of Galaxies as the Origin of Unidentified
G\lowercase{e}V Gamma-Ray Sources}

\author{Tomonori Totani$^1$ and Tetsu Kitayama$^2$}
\affil{ 
$^1$National Astronomical Observatory, Mitaka, Tokyo 181-8588, Japan
(e-mail: totani@th.nao.ac.jp) \\
$^2$Department of Physics, Tokyo Metropolitan University, Hachioji, 
Tokyo 192-0397, Japan (e-mail: tkita@phys.metro-u.ac.jp)}

\begin{abstract}
Over half of GeV gamma-ray sources observed by the EGRET experiment
have not yet been identified as known astronomical objects.  There is
an isotropic component of such unidentified sources, whose number is
about 60 in the whole sky. Here we calculate the expected number of
dynamically forming clusters of galaxies emitting gamma-rays by high
energy electrons accelerated in the shock wave when they form, in the
framework of the standard theory of structure formation.  We find that
a few tens of such forming clusters should be detectable by EGRET and
hence a considerable fraction of the isotropic unidentified sources
can be accounted for, if about 5\% of the shock energy is going into
electron acceleration. We argue that these clusters are very difficult
to detect in x-ray or optical surveys compared with the conventional
clusters, because of their extended angular size of about $\sim
1^\circ$. Hence they define a new population of ``gamma-ray
clusters''. If this hypothesis is true, the next
generation gamma-ray telescopes such as GLAST will detect more than a
few thousands of gamma-ray clusters. It would provide a new tracer of
dynamically evolving structures in the universe, in contrast to the
x-ray clusters as a tracer of hydrodynamically stabilized systems. We
also derive the strength of magnetic field required for the
extragalactic gamma-ray background by structure formation to extend
up to 100 GeV as observed, that is about $10^{-5}$ of the shock-heated
baryon energy density.
\end{abstract}

\keywords{cosmology: theory---diffuse radiation---large-scale structure
of the universe---galaxies: clusters: general---gamma rays: theory}

\section{Introduction}
The deepest image of the universe in the high energy gamma-ray band beyond
0.1 GeV has been obtained by the EGRET experiment (Hartman et al. 1999). 
Identified sources of the third EGRET catalog include the 
Large Magellanic Cloud, 
five pulsars, and 66 active galactic nuclei (AGNs) of the blazar class.
However, over half of the EGRET sources
have not yet been identified as known astronomical objects, and
their origin is one of the most interesting mysteries in astrophysics.
The distribution of these unidentified sources can be interpreted
as the sum of the Galactic component along the Galactic disc
($|b| \ltilde 40^\circ$)
and another isotropic (i.e., likely extragalactic) component 
(\"Ozel \& Thomson 1996; see also Fig. 2 of Mukherjee et al. 1995). 
Several candidates have been proposed
as the origin of the Galactic unidentified sources, including
molecular clouds, supernova remnants, massive stars, and radio-quiet
pulsars (see e.g., Gehrels \& Michelson 1999 and references therein).
However, almost no candidate has been proposed to explain the
extragalactic unidentified sources, except for undetected AGNs.  
Recently, Mirabal et al. (2000) have performed comprehensive
follow-up observations for one of the high-latitude unidentified EGRET sources
(3EG J1835+5918, $b=25^\circ$) in X-ray, optical, and radio wavebands.
They found that any known class of GeV gamma-ray sources including
blazars and pulsars cannot be the origin of 3EG J1835+5918, and 
it suggests that this source belongs to a new class of GeV gamma-ray emitters.

There are 19 unidentified sources with high galactic latitude of
$|b| > 45^\circ$ (about 60 in the whole sky), and 7 of them are noted as
`em' in the third EGRET catalog, i.e., possibly extended or multiple
sources that are inconsistent with single point sources.
Although this `em' designation is quite subjective and we should be 
careful in interpreting this result (see Hartman et al. 1999 for detail),
this may indicate that there is an extended and
extragalactic population in the unidentified EGRET sources. 
Recently Gehrels et al. (2000) chose `steady' unidentified 
sources from the third
EGRET catalog, to eliminate any source with high variability typical
of flaring AGNs. These steady sources are mostly distributed at
low and mid galactic latitude ($|b| \ltilde 40^\circ$)
and hence should be the Galactic origin,
but 7 sources are still located at $|b| > 45^\circ$, that may be
steady extragalactic sources whose number is $\sim 24$ in the whole sky.
Therefore, steady 
astronomical objects with an extended nature are worth being investigated
as a possible origin of unidentified EGRET sources.

It is widely believed that the observed structures in the universe
have been produced via gravitational instability. 
Currently the most successful theory of structure formation is the
cold dark matter (CDM) scenario, in which
the structures grow up hierarchically from small objects into larger
ones.  When an object collapses gravitationally and virializes, the
baryonic matter in the object is heated by shock waves up to the
virial temperature. Particles are expected to be accelerated to high
energy by shock acceleration, and accelerated electrons scatter the
photons of the cosmic microwave background radiation (CMB) to high
energy gamma-ray bands by the inverse-Compton mechanism. Existence
of such nonthermal electrons is inferred from radio and hard x-ray
observations for some clusters of galaxies (e.g., Fusco-Femiano et al. 
1999), although the origin of the nonthermal electrons is not yet clear.
It has also recently been argued that this radiation process in 
the intergalactic
medium may explain the diffuse extragalactic gamma-ray background
radiation (EGRB) observed in the EGRET range (Loeb \& Waxman 2000).
However, it is still highly speculative and difficult to test whether
this process is really the origin of the EGRB, since the contribution
by unresolved active galactic nuclei is also of the same order of
magnitudes (see, e.g., M\"ucke \& Pohl 2000 and references therein).

On the other hand, if the structure formation is actually an efficient
radiation process of gamma-rays, clusters of galaxies should be strong
emitters of gamma-rays when they dynamically form, and the
detectability of such forming clusters as discrete sources is of great
interest as a new probe of structure formation in the universe as well
as a test for the scenario proposed by Loeb \& Waxman (2000) for EGRB. In this
paper we make a theoretical estimate of the number and angular size of
such gamma-ray emitting clusters detectable by EGRET, based on the
standard theory of structure formation in the CDM universe.  We find a
few tens of such forming clusters should have already been detected by
EGRET. Detectability of such forming clusters in other wavebands such
as optical or x-ray bands will be discussed, in comparison with the
conventional clusters of galaxies identified in these wavebands. We
will also calculate the EGRB spectrum from structure formation and
derive a quantitative relation between the higher cut-off photon
energy and magnetic field strength.

Throughout this paper, we assume a CDM universe with the density
parameter $\Omega_0=0.3$, the cosmological constant
$\Omega_\Lambda=0.7$, the Hubble constant $h = H_0/(100
\mbox{km/s/Mpc})=0.7$, the baryon density parameter $\Omega_B = 0.015
h^{-2}$, and the density fluctuation amplitude $\sigma_8=1$. These
cosmological parameters are consistent with various observations
including those of the CMB fluctuations (e.g. de Bernardis et
al. 2000) and the abundance of x-ray clusters of galaxies (e.g. Eke, Cole \&
Frenk 1996; Kitayama \& Suto 1996b).

\section{Gamma-ray Luminosity and Flux from Forming Clusters}
\label{section:flux}
We first estimate the gamma-ray flux of a gravitationally bound object
of total mass $M$ that virializes at redshift $z$.  The typical
radius $r_{\rm vir}$, density $\rho_{\rm vir}$, circular velocity
$V_c$, and temperature $T_{\rm vir}$ of the object can be computed
from the spherical collapse model (Peebles 1980; Kitayama \& Suto
1996b), that is widely used in study of structure formation.  The
total gravitational energy given to the baryon gas in the forming
cluster is given by $E_{\rm baryon} \sim (3/4)(\Omega_B/\Omega_0) M
V_c^2$. It is reasonable to expect that a fraction $\xi_e \sim 0.05$
of this energy goes into accelerated electrons, since such a fraction
is inferred for acceleration of cosmic ray electrons in a supernova
remnant SN 1006 from x-ray and TeV observations (Koyama et al. 1995;
Tanimori et al. 1998) and consistent with the energetics among
cosmic-rays, turbulent motions, and supernova rate in our Galaxy. It
has also been suggested that the diffuse radio
and hard x-ray emissions observed in the Coma cluster 
(and possibly other several clusters) can be
attributed to nonthermal electrons with the electron
energy fraction of the same order (e.g., Fusco-Femiano et al. 1999).
Therefore, we use $\xi_e = 0.05$ to determine the normalization of
electron energy spectrum throughout this paper.

The maximum 
Lorentz factor of electrons is constrained by the competition of the Fermi
acceleration time and cooling time by inverse-Compton
(IC) scattering of CMB photons. The acceleration time is given by 
\begin{eqnarray}
t_{\rm acc} &\sim& \frac{r_L c}{V_s^2} \\ &=& 1.6\times 10^{-4} \gamma_e
B_{\mu G}^{-1} V_{s, 3}^{-2} \ \rm yr,
\end{eqnarray} 
where $r_L=m_e \gamma_e / (eB)$
is the Larmor radius of electrons, $\gamma_e$ the electron Lorentz
factor, $e$ the electron charge, $B_{\mu G} = B/(1\mu G)$ the magnetic
field, and $V_{s, 3} = V_s/(10^3 \rm \ km/s)$ the shock velocity that
is of the same order of magnitudes with the circular velocity of a
halo, $V_c$. On the other hand, the IC cooling time is 
\begin{eqnarray}
t_{IC} &=& \frac{\gamma_e m_e c^2}{(4/3) c \sigma_T U_{\rm CMB} \gamma_e^2} \\
&=& 2.3 \times 10^{12} \gamma_e^{-1} (1+z)^{-4} \rm \  yr, 
\end{eqnarray}
where $\sigma_T$ is the Thomson cross section and
$U_{\rm CMB} = 4.32 \times 10^{-13} (1+z)^4 \ \rm erg \ cm^{-3}$
is the CMB energy density.
Equating these expressions of $t_{\rm acc}$ and $t_{\rm IC}$, we have
the maximum value of $\gamma_e$ as
$\gamma_{e, \max} = 1.2 \times 10^8 (1+z)^{-2}
B_{\mu G}^{1/2} \ V_{s, 3}$. We assume the energy distribution of
accelerated electrons as a power-law with an exponential cut-off at
$\gamma_{e, \max}$, i.e., $dN_e/d\gamma_e \propto \gamma_e^{-\alpha}
\exp(-\gamma_e/\gamma_{e, \max})$, with the standard particle
acceleration index of $\alpha \sim 2$. As mentioned above,
the normalization of this spectrum is set 
by the equation
\begin{equation}
\int d\gamma_e m_e c^2 
\gamma_e \frac{dN_e}{d\gamma_e} = \xi_e E_{\rm baryon} \ ,
\end{equation}
with the parameter $\xi_e = 0.05$. The observed photon energy
$\epsilon_\gamma$ scattered by electrons is related to $\gamma_e$ as
$\epsilon_\gamma = (4/3)\gamma_e^2 \epsilon_{\rm CMB, 0}$, where
$\epsilon_{\rm CMB, 0} = 6.4 \times 10^{-4}$ eV is the mean photon
energy of the CMB at $z=0$.

The cooling time of electrons corresponding to photon energy 
$\epsilon_\gamma$ can be written 
as $t_{\rm IC} = 2.1 \times 10^{6} (\epsilon_\gamma/ {\rm GeV})^{-1/2} 
(1+z)^{-4}$ yr, and this should be compared with the time for the shock
wave to propagate the radius of the virialized halo, $t_{\rm shock}
\sim r_{\rm vir} / V_s = r_{\rm vir} / (4 V_c / 3)$. Here we have 
estimated the shock velocity as $V_s = (4/3) V_c$, that is a velocity
of a strong shock when a material is shocked by a supersonic piston
with a velocity of $V_c$, i.e., a typical bulk velocity of
material in a collapsed halo. By using the spherical collapse model
mentioned above to calculate $r_{\rm vir}$ and $V_c$ for the halo,
this time scale can be written as $t_{\rm shock} = (3/4)^{3/2} 
\pi^{-1/2} (G\rho_{\rm vir})^{-1/2} \sim 1.5 (1+z)^{-3/2}$ Gyr, 
that is essentially the dynamical time of the halo. Note that it depends only
on the redshift and not on the halo mass. From this argument the cooling 
time of electrons emitting gamma-rays above 0.1 GeV 
is always much shorter than the time scale $t_{\rm shock}$
during which the shock
is alive and a halo is an active gamma-ray emitter.
Hence the total number of gamma-rays emitted from a forming halo 
during the time $t_{\rm shock}$ is given as:
\begin{eqnarray}
\frac{dN_\gamma(\epsilon_\gamma; M, z)}{d\epsilon_\gamma} 
= \frac{m_e \gamma_e}{\epsilon_\gamma}
\frac{dN_e}{d\gamma_e} \frac{d\gamma_e}{d\epsilon_\gamma} \ ,
\end{eqnarray}
and the observed photon flux of gamma-rays during the shock propagation time is
\begin{eqnarray}
\frac{dF(\epsilon_\gamma; M, z)}{d\epsilon_\gamma}  
= \frac{(1+z)}{4 \pi d_L^2}
\frac{dN_\gamma}{d\epsilon_\gamma} \frac{1}{t_{\rm shock}} \ ,
\end{eqnarray}
where $d_L$ is the standard luminosity distance.

We introduce a parameter $\xi_B$ to determine the magnetic field,
which is the fraction of magnetic energy density in the total baryon
energy density of the halo, i.e., $B^2/(8\pi) = \xi_B E_{\rm baryon}
/(4\pi r_{\rm vir}^3/3)$.  Then by using the spherical collapse model 
again, the magnetic field of a cluster can be written as 
\begin{eqnarray}
B \sim 0.17 \left(\frac{\xi_B}{10^{-3}}\right)^{1/2} 
\left(\frac{M}{10^{15}M_\odot}\right)^{1/3} (1+z)^2
\ \mu \rm G \ . 
\end{eqnarray}
A magnetic field of $\sim$ 0.1--1 $\mu$G is often observed
in intracluster medium of
rich clusters (Kronberg 1994; Fusco-Femiano et al. 1999;
Rephaeli, Gruber, \& Blanco 1999),
and hence we use $\xi_B = 10^{-3}$ to be consistent
with the observations. It should be noted, however, that this parameter
is important only for the maximum photon energy of the gamma-ray spectrum
(well beyond 10--100 GeV with $\xi_B = 10^{-3}$, 
see \S \ref{section:EGRB}),
and the gamma-ray flux above 100 MeV measured by the EGRET
is almost insensitive to this uncertain parameter, unless
the particle index $\alpha$ significantly deviates from 
the standard value of 2.

It is interesting to apply the above model to nearby known clusters of
galaxies. For example, the Coma cluster has the total mass of $M \sim
10^{15} M_\odot$ and is located at $z = 0.023$. The flux with these
parameters becomes $F(>0.1 {\rm GeV}) \sim 6.5 \times 10^{-7} \ \rm
photons \ cm^{-2} sec^{-1}$, that is about 10 times brighter than the
observational upper limit on this cluster by EGRET, $4 \times 10^{-8}
\ \rm photons \ cm^{-2} sec^{-1}$ (Sreekumar et al. 1996).  This does
not mean, however, that our model is incorrect. Our model is relevant
for just-forming clusters of galaxies in which the violent shock
generated by gravitational collapse is still alive.  The Coma cluster
is thought to have formed more than a few dynamical times ago and is
now hydrodynamically stable after violent shock has
disappeared. Gamma-ray flux can then be much weaker than our
estimation. On the other hand, it suggests that there was an epoch
during which this cluster was a strong gamma-ray emitter, and also
that there may be other clusters visible by EGRET, that are just
dynamically forming and have not yet reached hydrostatic equilibrium.

\section{Expected Number of Gamma-Ray Clusters Detectable by EGRET}
The number of such forming clusters of galaxies with flux stronger than
$F$ can be calculated as
\begin{eqnarray}
N(>F) = \int dz \int_{M(z; \ F)}^\infty dM \ 
\frac{dV}{dz} \ R_{\rm form}(M, z) \ t_{\rm shock}  \ ,
\label{eq:lognf}
\end{eqnarray}
where $dV/dz$ is the comoving volume element of the universe, $R_{\rm
form}$ the formation rate of dark haloes (or clusters)
per unit mass, cosmic time, and comoving volume, and $M(z; \ F)$ the mass of
a cluster collapsing at redshift $z$ whose flux is $F$. Here we have taken into
account that clusters are active gamma-ray emitters only during the
time $t_{\rm shock}$.

\begin{figure*}[t]
\centerline{\epsfxsize=11.5cm \epsfbox{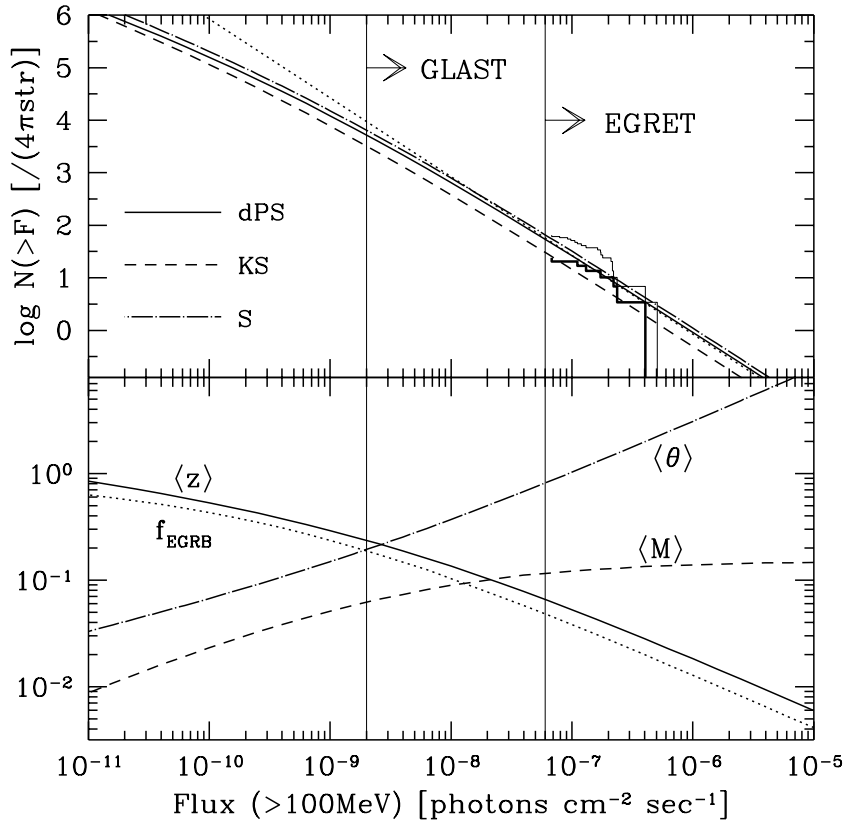}}
%\centerline{\epsfxsize=\linewidth \epsfbox{f1.ps}}
%  \begin{center}
%    \leavevmode\psfig{file=f1.ps,width=11cm}
%  \end{center}
\caption{The upper panel: the cumulative flux distribution of
gamma-ray emitting clusters of galaxies. The three curves are the
theoretical predictions using three different formulae for the
formation rate of dark haloes ($R_{\rm form}$), i.e., the derivative of the
Press-Schechter (PS) mass function (solid line), formation rate of Sasaki
(1994; dot-dashed line) and Kitayama \& Suto (1996a; dashed line).
The dotted line is the flux distribution of uniformly distributed
sources in the Euclidean space. The observed distribution of the
unidentified EGRET sources with $|b|>45^\circ$ is shown by the thick
(only the `em' sources) and thin (all sources including non-`em')
lines. The lower panel: the solid, dashed, and dot-dashed lines are
mean redshift, mass [$10^{16}M_\odot$], 
and angular radius [degree] of gamma-ray clusters brighter
than a given flux, respectively. 
The dotted line is the contribution of gamma-ray
clusters brighter than a given flux to the diffuse extragalactic
gamma-ray background at 100 MeV. The derivative of the
PS mass function is used.
 The sensitivity limits of the EGRET and GLAST experiments are shown
in the figure.  }
\label{fig:gamma-cluster}
\end{figure*}

There are several formulae to calculate the formation rate $R_{\rm
form}$ in the framework of the standard theory of structure formation.
As is well known, the Press \& Schechter (1974, PS; Peebles 1980) 
formalism provides a formula of mass function 
(i.e., number density of haloes as function of mass and redshift),
that is in reasonable agreement with $N$-body simulations.
Here we want the formation rate of haloes rather than
the mass function at a given epoch, because we need to calculate 
the number of collapsing objects experiencing shock at each epoch. A naive
prescription to obtain this quantity is to take a time derivative of
the PS mass function, $R_{dPS}$, although this is not exactly $R_{\rm
form}$ but rather interpreted as $R_{dPS} = R_{\rm form} - R_{\rm
dest}$, where $R_{\rm dest}$ is the rate of destruction of haloes by
merging into even larger structures. Consequently, $R_{\rm dPS}$
becomes negative at small mass scales where $R_{\rm dest}$ is
significant. As shown below, however, the number of objects visible by
EGRET is dominated by massive clusters forming in recent past $(z
\ll 1)$ that are the largest structures in the universe, and hence
$R_{\rm dest}$ is negligible.
Therefore it is a reasonable approximation to use $R_{dPS}$ in the mass range
where it is positive. Alternatively, there are several
extensions to the PS theory for computing $R_{\rm form}$ (e.g., Lacey
\& Cole 1993; Sasaki 1994; Kitayama \& Suto 1996a; Percival \& Miller
1999). In what follows we use $R_{dPS}$ and the formulae of Sasaki
(1994) and Kitayama \& Suto (1996a) to take account of theoretical
uncertainties in $R_{\rm form}$.

Figure \ref{fig:gamma-cluster}
shows the theoretically predicted $\log N$-$\log F$ of
forming clusters. At least a
few tens of clusters should be visible by the EGRET, and a
significant fraction of the isotropic unidentified EGRET sources can
be accounted for. 
It is also interesting to note that the predicted number is similar to
that of `em' isotropic unidentified EGRET sources, i.e., possibly
extended sources. The predicted number at the EGRET flux limit also
agrees with the number of `steady' unidentified 
sources with $|b|>45^\circ$ defined by Gehrels et al. (2000).
This result is robust against changes in the adopted
prescription for $R_{\rm form}$.
In the lower panel of Fig. \ref{fig:gamma-cluster}, 
we also show mean mass, redshift, and
angular radius $\theta_{\rm vir}$
corresponding to $r_{\rm vir}$ of such gamma-ray
clusters brighter than a given flux. These quantities are $M \sim
10^{15}M_\odot$, $z \sim 0.05$, and $\theta_{\rm vir}
\sim 1^\circ$ for clusters
above the EGRET sensitivity limit. Considering the EGRET angular
resolution, the typical radius of $\sim 1^\circ$ is
consistent with the fact that a significant fraction of isotropic
unidentified sources are indicated as possibly extended. 

It is predicted
that more than a few thousands of forming clusters will be detected by
future missions such as the GLAST (Gehrels \& Michelson 1999), and 
the improved angular resolution may reveal
the extended profile for nearby gamma-ray clusters with higher
statistical significance. Another important prediction is that GLAST
will observe the flattening of the $\log N$-$\log F$ curve due to the
cosmological effects, compared with the expectation of a uniform
source distribution in the Euclidean space (dotted line in the upper
panel of Fig. \ref{fig:gamma-cluster}).

\section{Extragalactic Gamma-ray Background}
\label{section:EGRB}

\begin{figure*}[t]
\centerline{\epsfxsize=11.5cm \epsfbox{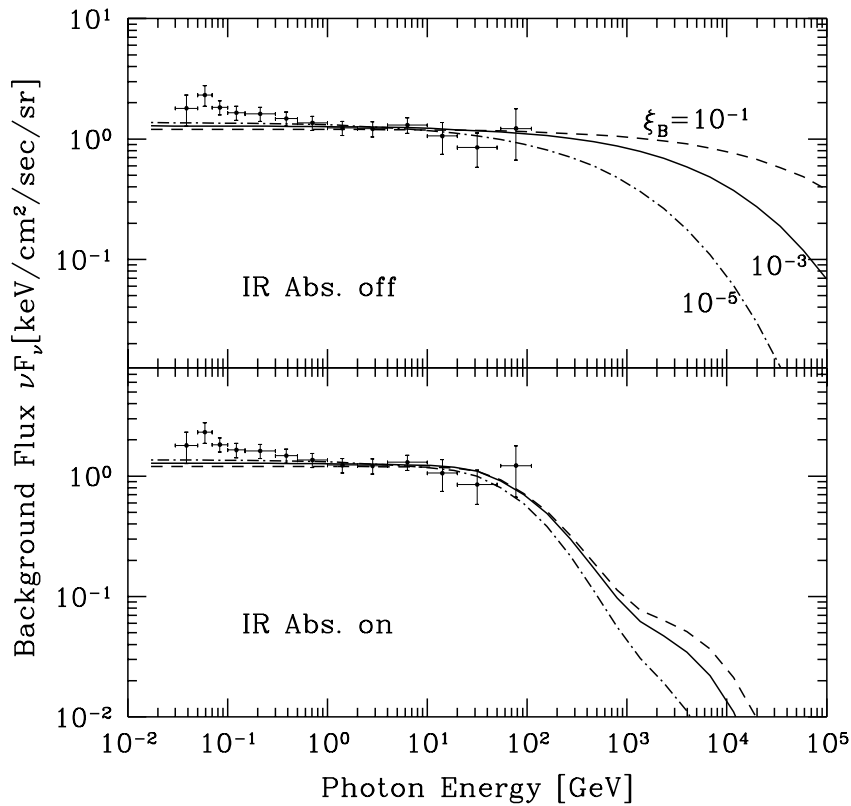}}
%  \begin{center}
%    \leavevmode\psfig{file=f2.ps,width=11cm}
%  \end{center}
\caption{The spectrum of the diffuse extragalactic gamma-ray background
radiation. The data is from Sreekumar et al. (1998).
The parameter $\xi_B$ is the fraction of magnetic energy in the
gravitational energy given to baryonic gas in a collapsed halo, 
with $\xi_B = 10^{-1}$
(dashed line), $10^{-3}$ (solid), and $10^{-5}$ (dot-dashed).
The absorption of
gamma-rays above $\sim$ 100 GeV by the pair-production
interaction with the cosmic infrared background is not taken
into account in the upper panel, while it is in the lower panel.
(The secondary gamma-rays reprocessed by the produced pairs 
are not taken into account in either of the panels, see text.)
}
\label{fig:EGRB-spec}
\end{figure*}

Our formulation also allows us to calculate the EGRB flux and spectrum as
\begin{eqnarray}
\frac{dn_\gamma}{d\epsilon_\gamma} = \int dz \int dM \
\frac{dN_\gamma(\epsilon_\gamma; M, z)}
{d\epsilon_\gamma} \ R_{\rm form}(M, z) \ \frac{dt}{dz} \ ,
\end{eqnarray}
where $(dn_\gamma/d\epsilon_\gamma)$ is the gamma-ray number density
that is related to the EGRB flux as $(dF/d\epsilon_\gamma) 
= c (dn_\gamma/d\epsilon_\gamma) /(4 \pi)$,
and $t$ is the cosmic time.
This flux becomes $\epsilon_\gamma^2 (dF/d\epsilon_\gamma) \sim 1 \rm keV \
cm^{-2} sec^{-1} sr^{-1}$ at 100 MeV, in good agreement with the
observation (Sreekumar et al. 1998) as well as the previous simpler estimation
assuming that the average temperature of baryons in the universe is
$\sim$ keV at present (Loeb \& Waxman 2000). In fact, we have checked
that the mass-averaged temperature of virialized haloes in the universe as a
function redshift, that is calculated by the PS theory, agrees within
a factor of 2 with a numerical simulation (Cen \& Ostriker 1999) on
which the previous EGRB estimate was based. This fact gives a
justification for the use of the PS theory to calculate the gamma-ray
emitting objects.  The dotted line in the lower panel of Fig.
\ref{fig:gamma-cluster} shows
the contribution to the EGRB at 100 MeV by objects brighter than a
given flux. We predict that GLAST will resolve about 20--30 \% of
the EGRB as discrete gamma-ray clusters, 
if structure formation is the major origin of the EGRB.

The strength of magnetic field in the shocked baryons 
is important for the question
whether the EGRB spectrum extends up to $\sim$ 100 GeV as observed.
In Fig. \ref{fig:EGRB-spec}
we show the EGRB spectrum with several values of $\xi_B$.
This result shows that the magnetic field strength corresponding to
$\xi_B \sim 10^{-5}$ of the baryon energy density is sufficient
for the EGRB to extend beyond 100 GeV. The magnetic field observed
in the intracluster gas ($\xi_B \sim 10^{-3}$) is much stronger than this, and 
it is also theoretically reasonable to expect that the turbulent
motion in collapsed objects amplifies the seed magnetic field
made by the battery mechanism well beyond $\xi_B \gtilde 10^{-5}$
within the dynamical time (Kulsrud et al. 1997). 
Therefore, physically reasonable magnetic field strength
can explain the extension of the EGRB spectrum beyond 100 GeV,
and it is likely that
a considerable fraction of gamma-rays above 100 GeV is absorbed by the 
interaction with the cosmic infrared background radiation
producing electron-positron pairs. The effect of intergalactic absorption
is shown in the lower panel of Fig \ref{fig:EGRB-spec}, 
using the optical depth of intergalactic pair-production 
presented in Totani (2000).
The absorbed TeV gamma-rays will be reprocessed into GeV gamma-rays
by the produced pairs, and distort the EGRB spectrum (e.g.,
Coppi \& Aharonian 1997). Although these secondary GeV gamma-rays are
not taken into account here and it is beyond the scope of this paper,
it is important to study how large is this spectral distortion in future work.

\section{Discussion}
We here discuss the expected properties of gamma-ray clusters of
galaxies. Perhaps the most natural question in this regard would be
``Are they already observed in other wavebands such as x-rays or
optical surveys?'' We have checked that
there is no statistically significant association of the ROSAT Brightest
Cluster Sample (RBCS, Ebeling et al. 1998), 
within the 95\% error circles of the unidentified sources with $|b|>30^\circ$
in the EGRET catalog. 
We have also checked the correlation with the clusters in
the revised Abell catalog (Abell, Corwin, \& Olowin 1989), and
no statistically significant associations are found, either.
However, in the following we argue that the gamma-ray
clusters proposed in this paper are very difficult to detect
in x-rays or optical bands compared with ordinary clusters
identified in these wavebands, and hence our scenario is not rejected by
these results.

\subsection{Detectability of gamma-ray clusters in X-rays}
\label{section:x-rays}
We first estimate the expected x-ray flux from gamma-ray clusters.
Baryonic gas in most clusters of galaxies observed in x-rays seems to
be in approximate hydrostatic equilibrium with the surface brightness
well fitted by a density profile, $\rho_{\rm gas}(r) \propto
[1+(r/r_c)^2]^{-1}$ (e.g., Sarazin 1988), where $r_c$ is the core
radius that is typically about $\sim 10$ times smaller than the virial
radius.  Since the x-ray emissivity is proportional to $\rho_{\rm
gas}^2$, the x-ray emission is strongly concentrated into the central
region.  Assuming the above density profile and the self-similar model
as described in Kitayama \& Suto (1996b), 
\footnote{Here we have assumed that
the core radius is proportional to the virial radius as $r_c = 0.21
(h/0.7)^{-1} {\rm Mpc } \times [r_{\rm vir}(M, z)/ r_{\rm
vir}(10^{15}M_\odot, 0)]$ with $r_{\rm vir}(10^{15}M_\odot, 0)$ = 2.6Mpc, 
where the normalization is chosen to match
the local observations (Abramopoulos \& Ku 1983; Jones \& Forman 1984).}  
a typical gamma-ray cluster
detectable by the EGRET with $M \sim 10^{15} M_\odot$ and $z \sim
0.05$ would have the x-ray flux $2.4 \times 10^{-11} \rm \ erg \
cm^{-2} \ s^{-1}$ in 0.1--2.4 keV.

The inverse-Compton flux is also expected to be comparable with the
thermal emission. By equating $t_{IC}$ and $t_{\rm shock}$ in \S
\ref{section:flux},
we get the cooling photon energy $\epsilon_{\rm \gamma, cool}
= 2.0 (1+z)^{-5}$ keV, below which the electron cooling time is
longer than the dynamical time. Then the IC spectrum 
extends down to around x-ray band with 
$dN_\gamma/d\epsilon_\gamma \propto \epsilon_\gamma^{-2}$,
while it becomes harder at wavelengths longer than x-rays with
$dN_\gamma/d\epsilon_\gamma \propto \epsilon_\gamma^{-1.5}$.
If the gamma-ray flux at 100 MeV
is $\sim 10^{-7} \rm \ photons \ cm^{-2} s^{-1}$
that is the EGRET threshold, the IC x-ray flux ($\nu F_\nu$) is
$\sim 1.6 \times 10^{-11} \rm \ erg \ cm^{-2} \ s^{-1}$.
Therefore, the thermal and IC
fluxes are well above the flux limit $\sim 4 \times 10^{-12} \rm
\ erg \ cm^{-2} \ s^{-1}$ of the RBCS.

However, it takes nearly the dynamical time for the cluster gas to
reach hydrostatic equilibrium after the collapse, and gamma-rays from
the shock generated by the gravitational collapse are radiated away within that
period. Then it is likely that the density profile of gamma-ray
emitting clusters is more irregular and extended than 
ordinary x-ray clusters.  In
fact, if the unidentified `em' sources in the EGRET catalog
are actually extended, they must have typical angular size of 
about degree, from the source location accuracy of the EGRET.
As we have shown, angular size of about 1$^\circ$ is
theoretically reasonable if the emission is extended to the virial
radius.  When the density profile is not concentrated into the central
region but rather constant within the virial radius, the x-ray luminosity
becomes lower than the self-similar model
by a factor of $\sim 3.7$ because of the lower central
density.

Furthermore, the surface brightness of such loose
clusters should be drastically dimmer than ordinary x-ray clusters.  
In the self-similar model with $r_{\rm vir} \gg r_c$, the core gas density is
$\rho_{\rm gas, c} \sim (1/3)(r_{\rm vir}/r_{c})^2 
\rho_{\rm gas, vir} \sim 50 \rho_{\rm gas, vir}$, where
$\rho_{\rm gas, vir} = (\Omega_B/\Omega_0) \rho_{\rm vir}$ is the
virial gas density that is the average gas density within $r_{\rm vir}$. 
On the other hand, if the gas density
profile of gamma-ray clusters is roughly constant at 
$\rho_{\rm gas, vir}$ out to $r_{\rm vir}$, 
the x-ray surface brightness of such a loose cluster 
is dimmer than the central surface brightness of the self-similar model
by a factor of 
$\sim (r_c/r_{\rm vir})(\rho_{\rm gas, c}/
\rho_{\rm gas, vir})^2 \sim 200$, since the
x-ray emissivity is proportional to $\rho_{\rm gas}^2$.
It crucially affects the
detectability of x-rays from gamma-ray clusters. The detectability of
x-rays should be described by the signal-to-noise ratio $(S/N)$
against the x-ray background flux.  The noise level is proportional to
(image area)$^{1/2}$, and hence $S/N \propto F/r$, where $F$ and $r$
are the flux and the image radius, respectively.  We have compared the
value of $F/r$ of the extended gamma-ray clusters detectable by the
EGRET and those of the clusters in the RBCS.  We found that the $F/r$
of gamma-ray clusters is by a factor of 3 smaller than the minimum
$F/r$ of the RBCS clusters.  The absence of association between the
RBCS and the EGRET sources is therefore not in contradiction to our
scenario. On the other hand,
deeper observation of candidate gamma-ray clusters by 
Newton, for example, might detect the x-ray emission extended to about
$1^\circ$ with the flux estimated above, that would provide a clear
test of our scenario.  Such x-ray emission should reflect the
structure of shocks in dynamically forming clusters, and imaging study
is of great interest.

\subsection{Detectability in the Optical Surveys}
Here we again emphasize that the gamma-ray clusters are expected to be
more extended than clusters that have already stabilized.  
It is known that the surface density profile of galaxies in a cluster can well
be described by the King profile, $\sigma(r) \propto [1 +
(r/r_c)^2]^{-1}$ with the core radius of $\sim$ 100 kpc that is
comparable with the core radius of x-ray profile (e.g., Adami et al. 1998).
If we assume a roughly constant surface density out to $\sim r_{\rm
vir}$ rather than the King profile
for gamma-ray clusters, the
average surface density $\sigma_{\rm av} \sim N_{\rm gal}/(\pi r_{\rm vir}^2)$
is lower than the central surface density of the King profile
$\sigma_c \sim N_{\rm gal}/[2 \pi r_c^2 \ln(r_{\rm vir}/r_c)]$,
by a factor of $\sigma_c/\sigma_{\rm av}
\sim [2 \ln(r_{\rm vir}/r_c)]^{-1}(r_{\rm vir}/r_c)^2
\sim 30$. Here $N_{\rm gal}$ is the total number of galaxies within
$r_{\rm vir}$. This dimming factor is not so significant as that for the
x-ray surface brightness, but that should make the optical identification
very difficult because of the contamination by foreground and/or background
field galaxies. Therefore, we consider that no statistically
significant association with the known optically identified
clusters does not immediately reject our scenario. Instead, it is
necessary to study in the future the correlation between the EGRET
sources and galaxy catalogs taking into account the possibility that
the gamma-ray clusters are considerably extended. Search in optical
bands has an advantage over the search in x-rays, in a sense that the dimming
of surface number density compared with ordinary clusters
is less severe than x-rays whose emissivity is
proportional to $\rho_{\rm gas}^2$. The typical density
of such loose clusters is close to the virial density, that is about
a few hundreds times higher than the mean density of the universe.

We have also noticed that there are a considerable
number of `em' sources in the EGRET sources identified as AGNs.
If they were actually extended sources,
it might be speculated that some of them are also gamma-ray clusters
including an AGN as a member galaxy. Time variability of these sources would be
an important test to check this possibility.

\subsection{On the recent follow-up observations for 3EG J1835+5918}
Recent follow-up observations by Mirabal
et al. (2000) for one of the high-latitude unidentified EGRET sources
(3EG J1835+5918) have found a diffuse x-ray emission from an uncatalogued
cluster of galaxies at $z=0.102$. Although this cluster is outside the 99\%
error ellipse of 3EG J1835+5918 whose radius is 12$'$, 
the separation between the centers of the x-ray cluster and
3EG J1835+5918 is about 0.65$^\circ$, that is within our expectation of
the typical angular radius of gamma-ray clusters detectable by the
EGRET, $\sim 1^\circ$. As discussed above, x-ray emission is expected
from a region where the intracluster gas reached hydrodynamical
equillibrium, while gamma-rays are emitted from a region still
hydrodynamically unstable. Therefore, it is not surprising that
the positions of x-ray and gamma-ray emissions are different unless
the separation is well beyond the virial radius of $\sim 1^\circ$.

3EG J1835+5918 is not an `em' source, and this source
may not be an extended source. The radius of the 99\% confidence ellipse,
$0.2^\circ$, is considerably smaller than
the expected angular size of gamma-ray clusters. However, high energy electrons
emitting GeV gamma-rays have very short life time compared with the
shock propagation time ($t_{\rm IC} \sim 10^{-3} t_{\rm shock}$), 
and gamma-ray emitting region may be very clumpy
in a cluster. (On the other hand, x-ray emitting electrons have a
cooling time comparable with or
longer than $t_{\rm shock}$ for IC and thermal radiations, 
and hence x-ray emitting region
should be much less clumpy and
extended with the size $\sim \theta_{\rm vir}$, as discussed
in \S \ref{section:x-rays}.) Therefore, it is possible that the 
gamma-ray size of the 3EG J1835+5918 is considerably smaller than the
physical size of a whole forming cluster. This consideration also
suggests a possibility that some of gamma-ray clusters may be
observed as multiple sources within $\sim \theta_{\rm vir}$, 
that may be revealed by future gamma-ray missions.

One of the characteristics that the source of 3EG J1835+5918 must have
is very weak radio flux that is at least two orders of magnitudes
fainter than any of the securely identified EGRET blazars (Mirabal et
al. 2000). The spectrum of blazars is well understood by the two
components of radiation by the same population of nonthermal
electrons, i.e., synchrotron radiation in radio, optical, and X-ray
bands, and inverse-Compton radiation in GeV and TeV gamma-ray bands
(e.g., Inoue \& Takahara 1996; Kataoka et al. 1999). The ratio of
luminosities by the two processes is, as is well-known, given by the
ratio of magnetic energy density to the target photon energy density,
that is typically of order unity for blazars. On the other hand, this
ratio is $U_B/U_{\rm CMB} \sim 2.7 \times 10^{-3} (\xi_B/10^{-3})
(M/10^{15}M_\odot)^{2/3}$ for gamma-ray clusters, 
by using the expression of $B$ given in \S
\ref{section:flux}.  Therefore, the $U_B/U_{\rm CMB}$ ratio is generally 
much smaller than the unity, and very weak radio
flux compared with identified blazars can be reasonably
explained. 

Based on the above arguments,
we suggest that the uncatalogued x-ray cluster near
3EG J1835+5918 may be a gamma-ray cluster proposed in this paper,
and further observations for this cluster and the surrounding region
are very important. Our model predicts that a cluster emitting a flux of
$\sim 6.06 \times 10^{-7} \ \rm photons \ cm^{-2} s^{-1}$ above 100 MeV
(the flux of 3EG J1835+5918, Hartman et al. 1999) at $z$ = 0.102
should have a total mass of $\sim 7 \times 10^{15} M_\odot$ and
$r_{\rm vir} \sim 5$ Mpc ($\theta_{\rm vir} \sim 0.7^\circ$).

\section{Conclusions}
In this paper we have proposed a new candidate of unidentified EGRET
sources: gamma-ray clusters that are just dynamically forming and emit
gamma-rays due to inverse-Compton scattering of CMB photons by
shock-accelerated electrons. Based on the standard theory of structure
formation and assuming the injection of $\sim$5\% 
of shock energy at the formation into nonthermal electrons, 
we have shown that a few tens of such clusters should have
already been detected by EGERT, and a significant fraction of
the isotropic component of unidentified EGRET sources can be accounted
for. Such gamma-ray clusters are expected to be very extended;
the x-ray surface brightness and surface number density of galaxies
could be lower than those of
ordinary clusters by a factor of $\sim 200$ and $\sim 30$, respectively.
Therefore it should
have been very difficult to detect gamma-ray clusters in the past
x-ray or optical surveys, and our scenario is in accord with apparent
no-associations between unidentified EGRET sources and x-ray or
optical clusters.

It will be of great significance to perform x-ray or optical
observations to search for such loose clusters of galaxies in the
regions of high-latitude unidentified EGRET sources. The future
gamma-ray projects such as GLAST will also provide a direct test of our
scenario. If our scenario is true, a new population of ``gamma-ray
clusters'' will provide us in the future a
new probe of dynamically evolving structures in the universe
that cannot be traced by x-ray or optical clusters of galaxies.

We would like to thank S. Inoue and T. Naito for useful discussions.
TT has partially been supported by the Grant-in-Aid for the
Scientific Research Fund (No. 12047233) of the Ministry of Education, Science,
and Culture of Japan. TK has been 
supported in part by the Research Fellowships 
of the Japan Society for the Promotion of Science for Young Scientists.

\end{document}